# Hydrogenless Superluminous Supernova PTF12dam in the Model of an Explosion inside an Extended Envelope

**P. V. Baklanov**[1*], **E. I. Sorokina**[1,2**], **and S. I. Blinnikov**[1,2,3***]

[1]*Institute for Theoretical and Experimental Physics,*
*ul. Bol'shaya Cheremushkinskaya 25, Moscow, 117218 Russia*

[2]*Sternberg Astronomical Institute, Moscow State University,*
*Universitetskii pr. 13, Moscow, 119992 Russia*

[3]*Kavli IPMU (WPI), the University of Tokyo, 5-1-5 Kashiwanoha, Kashiwa, Chiba 277-8583, Japan*

**Abstract**—A model of a supernova explosion inside a dense extended hydrogenless envelope is proposed to explain the properties of the light curve for one of the superluminous supernovae PTF12dam. It is argued in the literature that the flux of this supernova rises too fast to be explained by the explosion model due to the instability associated with the electron–positron pair production (pair-instability supernova, PISNe), but it is well described by the models with energy input by a magnetar. We show that the PTF12dam-type supernovae can be explained without a magnetar in a model with a radiative shock in a dense circumstellar envelope that does not require an excessively large explosion energy.

## INTRODUCTION

Rare superluminous supernovae (SLSNe) whose luminosity at maximum light can exceed $10^{44}$ erg s$^{-1}$, which is greater than the typical values for core-collapse supernovae (CCSNe), $\sim 10^{42}$ erg s$^{-1}$, by two orders of magnitude, are encountered among the supernovae (SNe). Modern astronomy allows SLSNe to be observed at redshifts $z > 1$ (Cooke et al. 2012) owing to their high luminosity, which makes them very valuable sources of information at cosmological distances.

According to the standard classification of SNe, SLSNe were divided into two subclasses: with hydrogen lines in the spectrum (SLSNe-II) and hydrogen-poor ones (SLSNe-I) (Gal-Yam 2012; Quimby 2013). In addition, helium lines have never been observed in SLSNe-I; therefore, these SNe are more accurately classified as type Ic ones. In this paper, we will investigate the latter type of objects.

There is no universally accepted model for SLSNe-I. Several alternative scenarios are considered: the explosion of a star with an initial mass greater than 140 $M_\odot$ (PISN) with the production of an enormous amount of radioactive nickel (several or even tens of solar masses) followed by the envelope's heating from radioactive decays according to the chain $^{56}$Ni $\Rightarrow$ $^{56}$Co $\Rightarrow$ $^{56}$Fe; the envelope's heating through the reprocessing of the rotational energy released by a spinning-down millisecond magnetar; and the reprocessing of the kinetic energy of the shock into radiation produced when the SN ejecta interact with the surrounding extended dense envelope. So far there are no convincing arguments for a specific model.

Some of the SLSNe Ic exhibit a slope on the decline of their light curves that is characteristic of radioactive cobalt decay. Such tails of the light curves can be observed for several hundred days. One of the first such SLSNe Ic was SN 2007bi. The nickel mass capable of providing the observed flux on the decline of the light curve was determined by modeling its light curve and spectrum. In the PISN model, $M_{\mathrm{Ni}} = 3-7\ M_\odot$; an enormous explosion energy, $E = (0.8-1.3) \times 10^{53}$ erg, must be produced in this case (Gal-Yam et al. 2009; Kozyreva et al. 2014). Below, we will also use the energy unit foe: 1 foe $\equiv 10^{51}$ erg. Thus, SLSNe Ic with the PISN mechanism must explode with an energy of $\sim$100 foe. Moriya et al. (2010) showed that the observed light curves were reproduced both in the PISN model with $M_{\mathrm{ej}} =$

*E-mail: baklanovp@gmail.com, Petr.Baklanov@itep.ru
**E-mail: sorokina@sai.msu.su
***E-mail: Sergei.Blinnikov@itep.ru

121 $M_\odot$ and in the CCSN model (core-collapse SNe) with a synthesized $^{56}$Ni mass near $M_{Ni} = 6.1$ $M_\odot$, but a very large explosion energy was also required, $E = 3.6 \times 10^{52}$ erg= 36 foe.

Another SLSN Ic, PTF12dam, whose light curve after the peak is very similar to the light curve of SN 2007bi, was discovered not so long ago. In contrast to the latter, PTF12dam was discovered at the rise phase of the light curve; therefore, it is well known how its flux rose to the peak. This imposes additional constraints on the possible explosion models. Nicholl et al. (2013) showed that the PISN mechanism led to an excessively long rise time of the light curve for PTF12dam and proposed instead a model with energy input by a magnetar that did not require extremely high energies. It should be emphasized that Nicholl et al. (2013) did not construct a self-consistent model with allowance made for the interaction of the magnetar's radiation with the ejecta and did not consider the transfer of photons in the inner and outer SN layers. The model proposed by them is basically a roughly evaluative one and shows that the rise time and the peak luminosity could be explained at a sufficiently large initial angular momentum of a magnetar with a high magnetic field.

In this paper, we propose not an evaluative but a detailed radiation-hydrodynamics model of this interesting object without invoking the energetics of a magnetar. Based on our numerical calculations of the radiative transfer in the entire SN volume, we show that very powerful and long SLSNe Ic like PTF12dam can be explained in terms of the model of a supernova explosion inside an extended envelope (Grasberg and Nadyozhin 1986; Chugai et al. 2004; Woosley et al. 2007; Moriya et al. 2013; Baklanov et al. 2013). The possibility of the formation of such envelopes is justified, for example, in Woosley et al. (2007) and Moriya and Langer (2014). An extended envelope efficiently reprocesses the kinetic energy of the radiative shock propagating through it into radiation from several months to $\sim$1 year. The rise time of the light curve to the peak depends on the chemical composition and structure of the envelope. We find a model for which the rise time and the decline rate of the light curve for PTF12dam correspond to the observations. In this case, we are able to reproduce not only the bolometric luminosity but also the fluxes in the main filters and the behavior of the color temperature.

## OBSERVATIONS OF SN PTF12dam

SN PTF12dam was discovered on May 23, 2012, at the Palomar Transient Factory (Quimby 2012). There were no signatures of H and He lines in the spectra of PTF12dam, which allowed it to be classified as SN Ic. The measured redshift to the host galaxy is $z = 0.107$. The discovery of this SN is valuable in that the fluxes were observed before maximum light was reached. The fast rise of the light curves to the peak excludes the PISN models with a large amount of $^{56}$Ni, which have a long radiative diffusion time scale in the envelope and whose light curves slowly rise to the peak (Nicholl 2013). Observations reveal a considerable broadening of spectral lines corresponding to characteristic velocities $\approx 10^4$ km s$^{-1}$ during the entire period of observations of this SLSN. High velocities at the peak luminosity are generally a characteristic feature of all SLSNe Ic, but they are an order of magnitude lower, $\sim 10^3$ km s$^{-1}$, in most cases.

## MODELING

### General Properties of the Models

The presupernova models were constructed by a nonevolutionary method described previously (Baklanov et al. 2005; Blinnikov and Sorokina 2010). A quasi-polytrope in hydrostatic equilibrium was constructed in the inner regions. The temperature in this region is related to the density as $T \propto \rho^{0.31}$. After an artificial explosion at the center, we call this region "ejecta." Its mass and radius are $M_{ej}$ and $R_{ej}$, respectively.

We surround the ejecta by a dense expanding envelope whose origin is of no great importance for our modeling. The envelope could be formed, for example, by a previous explosion (or several explosions) in the model with pulsational pair instability (Woosley et al. 2007), an intense outflow of the presupernova within several months or years before its explosion, single or multiple mergers of stars. The velocity profile in the envelope is chosen to be similar to that obtained in the evolution calculations by Woosley et al. (2007); more specifically, the velocity in the bulk of the envelope is considerably lower than that of the ejecta and can increase considerably in the outer layers (see the model profiles in Fig. 5). A shock is produced at the interface between the ejecta and the circumstellar envelope, where the kinetic energy of the ejecta is efficiently converted into the thermal motion of particles and into radiation.

The density distribution in an extended envelope is specified by a power-law profile $\rho \propto r^{-p}$. An example of the density profile is shown in Fig. 1. We chose $p = 1.8$ for our model, following one of the most suitable models from Sorokina et al. (2015). The mass and radius of the envelope are designated as $M_{env}$ and $R_{env}$. The elemental abundances in the entire model were homogeneous. We used a carbon−oxygen model without hydrogen with a variable carbon-to-oxygen ratio and a helium model. Heavy elements accounted for about 2% of the total mass with abundances corresponding to the solar composition.

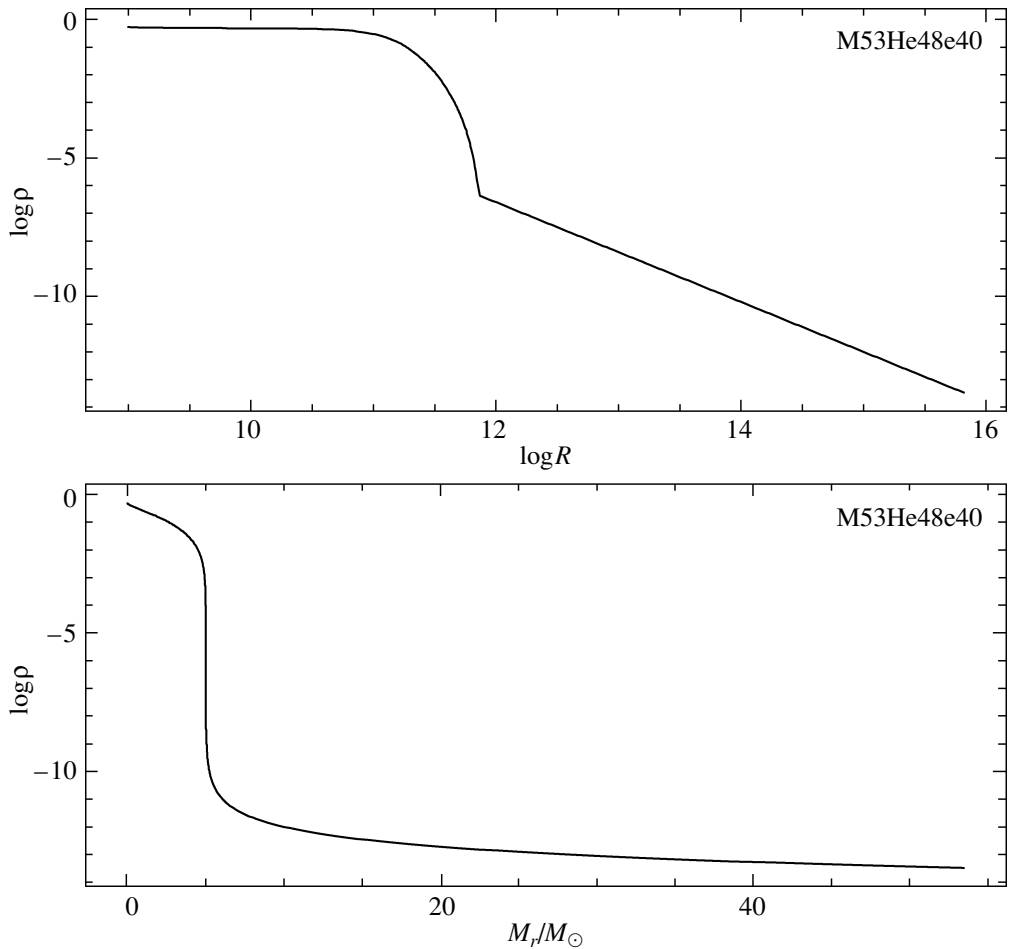

**Fig. 1.** Radial density profile. The upper panel: the logarithm of radius in centimeters is along the horizontal axis. The lower panel: density versus $M_r$, the mass within radius $r$, i.e., versus Lagrangian mass coordinate.

### Choosing a Specific Model

The main reason why Nicholl et al. (2013) rejected the PISN model for PTF12dam was an excessively long rise time of the model light curve. In this paper, we show that it is possible to choose such ejecta and envelope parameters in the ejecta–circumstellar medium interaction model that the rise time will correspond to the observations. In addition, we, of course, chose a model with a fading time and a luminosity at the peaks in different bands corresponding to their observed values. We did not seek to achieve a close coincidence with the observations, because in our modeling we used many of the numerical approximations, such as a spherical symmetry, an

LTE approximation for the ionization and population of atomic levels, and others. Since more realistic approximations can slightly change the results, it was important for us to show that all of the important light-curve parameters were reproduced in the model with interaction.

It seems interesting to compare the carbon–oxygen and helium models. The details of this study can be found in Sorokina et al. (2015). Here, we will describe the properties of the models with a dense extended envelope only briefly. At the time of explosion, the envelope is cold and optically transparent for most wavelengths. The generated radiative shock gradually heats up the envelope and it becomes optically thick. As the envelope heats up, the photosphere, which was near the shock shortly after the explosion, moves toward the outermost layers of the envelope by the time of peak luminosity. Thus, the rate of flux rise to the peak depends on the velocity with which the photosphere moves outward. This velocity is different for different chemical compositions of the gas, because it depends on the temperature at which the material becomes opaque. Helium needs to be heated to a higher temperature than does a carbon–oxygen mixture in order that its opacity increases considerably. That is why the rise phase of the light curve in the model with a helium envelope turns out to be appreciably longer than that in the model with a CO envelope.

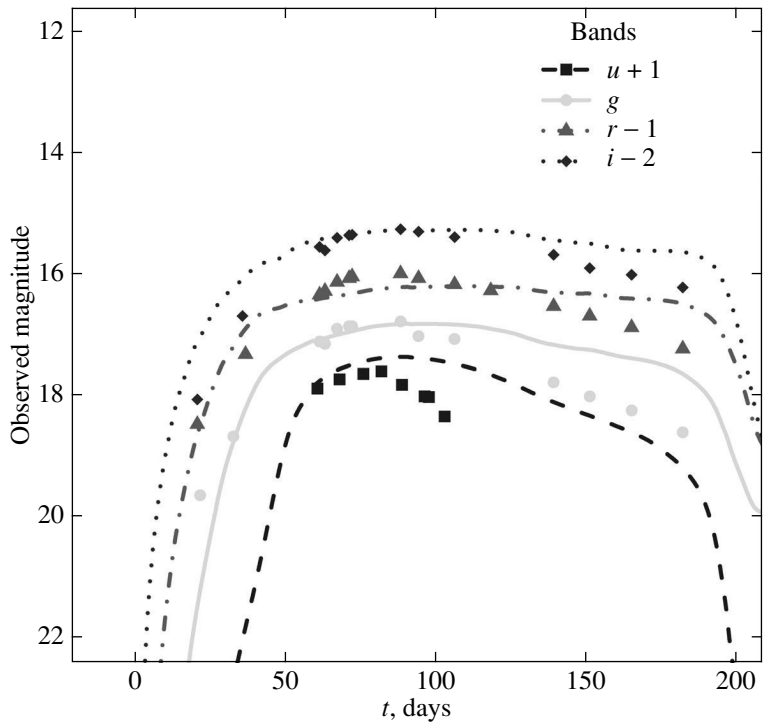

**Fig. 2.** *ugri* light curves for the `M53He48e40` model. The time in days after the explosion is along the horizontal axis, and the observed magnitude is along the vertical axis.

Our previous studies of SN 2010gx with a much narrower light curve (Blinnikov and Sorokina 2010) show that the rise time of the light curves for a CO model with a mass of the envelope ∼5 $M_\odot$, its radius ∼$10^{16}$ cm, an ejecta mass of 0.2 $M_\odot$, and an explosion energy of 2–4 foe is only 15–20 days, while the luminosity in the case of PTF12dam rises for almost two months. To increase the envelope heating time to such values, its mass needs to be increased considerably, to 50–100 $M_\odot$. Thus, we obtain a fairly exotic model in which hydrogen and helium must be lost by the star long before its explosion for them to be no longer visible in the SLSN spectra, while several tens of solar masses of carbon and oxygen must be lost shortly before its explosion. However, it should be remembered that SLSNe are also very rare and, therefore, may well be explained by the exotic model. An advantage of such a model over other models is that it is energetically economical: an explosion energy that exceeds only slightly the standard 1 foe for classical SNe is required to explain such an intense emission, while, for example, the PISN mechanism may require an explosion energy of several tens of foe.

In view of the properties of helium models described above, we nevertheless decided to consider precisely such a model as the main one for PTF12dam, despite the absence of helium signatures in the observed spectrum, because a lower helium mass is required to explain the observed rise time of the light curve and it is easier to produce a larger helium circumstellar envelope than a carbon–oxygen one with the same mass. Observers (Quimby et al. 2011) accept that helium is generally difficult to detect, although they believe that if it were present in the envelope in quantities of ten or more solar masses, then some of its traces would still be observed.

Thus, we chose a helium model consisting of 5 $M_\odot$ ejecta with an initial radius $R_{ej} = 10\ R_\odot$ expanding into a circumstellar envelope with a mass of 48 $M_\odot$ and a radius $R_{env} = 10^{16}$ cm as the best one.

The SN explosion was simulated by the release of $E_{exp} = 4 \times 10^{51}$ erg = 4 foe in the form of a "thermal bomb" in the innermost region of the ejecta. Figure 2 presents the light curves corresponding to our model and those obtained from the observations for broad-band ugri photometry in the period from the explosion to 250 days. It follows from our comparison with the observations that the SN explosion began at the epoch $MJD = 56\,010$, i.e., seven days earlier than for the magnetar model from Nicholl et al. (2013).

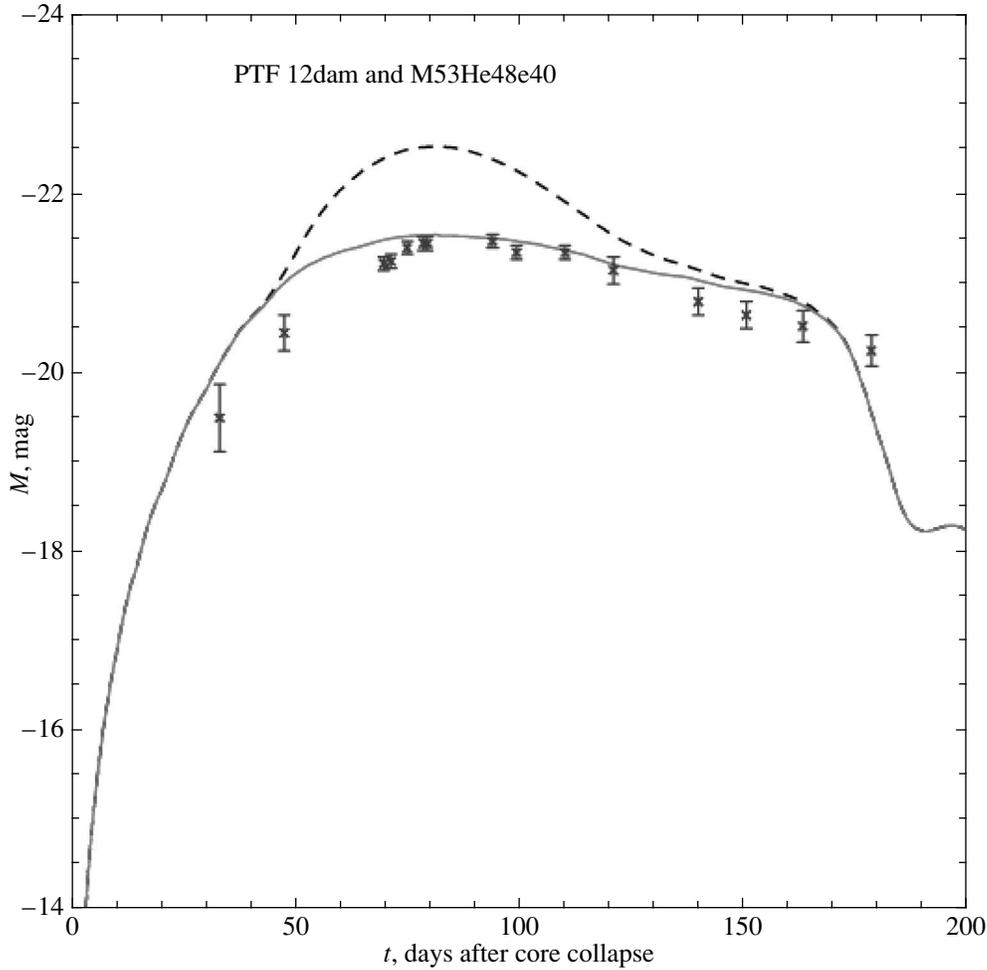

**Fig. 3.** Bolometric light curve for the helium `M53He48e40` model (dashes). The solid line indicates the light curve integrated from the ultraviolet part of the spectrum to its infrared part. The time in days after the explosion is along the horizontal axis, and the absolute magnitude is along the vertical axis. The observational data (Chen et al. 2014) are indicated by the crosses with error bars.

*Bolometric Light Curve*

The observational bolometric data were obtained by adding the contributions from the ultraviolet (UV), optical, and near-infrared (NIR) spectral ranges. Chen et al. (2014) described in detail the procedure to determine the bolometric light curve and pointed out the causes of some discrepancy between the observational data and the data from Nicholl et al. (2013). Since there are no continuous series of observations in all bands from UV to NIR for SN PTF12dam, observers have to resort to various kinds of tricks to reconstruct the missed data. In their analysis, they try primarily to use the real, measured fluxes. Therefore, if there were no data in optical bands in some period of observations, then these were found by a linear interpolation between the observed points closest to them. When the interpolation was impossible, for example, for the UV and NIR bands, the data for similar SNe in the coincident period were used. In the long run, if this did not help either, then a blackbody approximation was applied for those parts of the spectrum where the data were lacking.

Our modeled bolometric light curve and the observational data are presented in Fig. 3. The hot radiation that contributes significantly to the total flux integrated in the range from 1 Å to 5 $\mu$m (dashed line) dominates near the peak. For comparison with the observations, we calculated the flux integrated between 3000 Å and 4.5 μm (solid line).

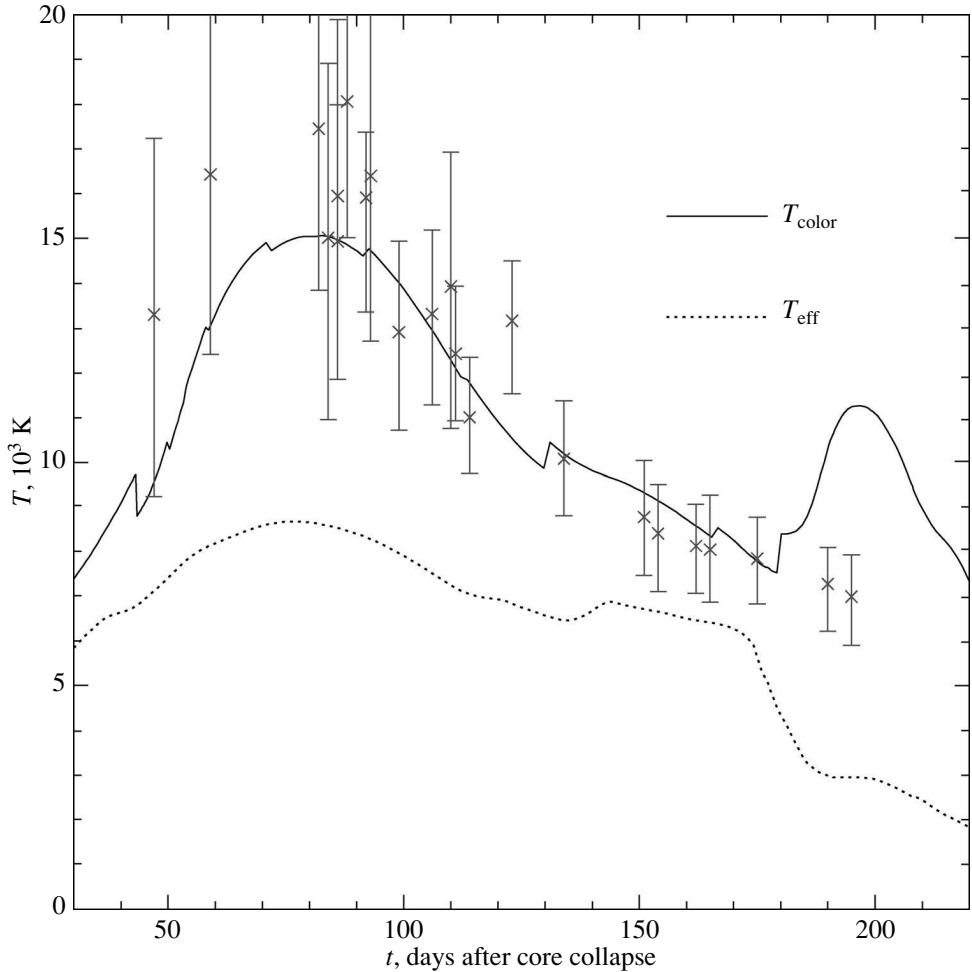

**Fig. 4.** Time variations of the color and effective temperatures determined from the spectrum for the `M53He48e40` model. The crosses indicate the observational data. The time in days is along the horizontal axis, and the temperature is along the vertical axis.

We obtained data on the effective temperature from Fig. 4 of Nicholl et al. (2013).

In the paper by Nicholl et al. (2013), the points along the time axis are measured from the peak of the light curve (apparently in the $R$ band). For our models, we supposed that the peak occurred on day 100. The shape of the curves agrees satisfactorily with the observations (see Fig. 4), but a higher temperature and, hence, less massive SNe at the same explosion energy are required.

Generally speaking, it is unclear what Nicholl et al. (2013) mean by $T_{\rm eff}$. It is very likely that it been obtained by fitting the observed spectra by blackbody radiation, i.e., what we call the color temperature $T_{\rm color}$ (the solid line in Fig. 4). The effective temperature $T_{\rm eff}$ determined via the integrated intensity

$\sigma T^4 = \pi J = \pi \int J_\nu d\nu$ passes well below $T_{\rm color}$ (the dashed line in Fig. 4). $T_{\rm color}$ is seen to agree with the observations within the error limits. The difference between $T_{\rm color}$ and $T_{\rm eff}$ shows that the radiation is diluted significantly. It is worth noting that the agreement with the observations (the position of the $T_{\rm color}$ peak) for our model is considerably better than that for the magnetar model (cf. Fig. 4 of Nicholl 2013). The surge in $T_{\rm color}$ near day 200 is related to the shock breakout. The absence of such a feature in the observations most likely suggests that the outer radius of the envelope is appreciably larger than is adopted in our calculations. However, this by no means can be an argument against the applicability of our model in principle.

The question about the velocity of the outer layers is to be also investigated. As can be seen from Fig. 5, the velocities reach several thousand kilometers per second at the photospheric level in our models, while observers estimate the velocities (though with a great uncertainty) to be $10\,000$ km s$^{-1}$. Here, the initial models are to be fitted further, but there should be no fundamental difficulties. The best strategy would be the construction of detailed evolutionary models for preceding explosions similar to those considered in Woosley et al. (2007). Interestingly, the ejecta from a preceding explosion in the hydrogen models from Woosley et al. (2007) had long remained neutral and transparent; therefore, the velocity at the photospheric level was only a few hundred kilometers per second, while a velocity of the outer layers was $4000$ km s$^{-1}$. In our hydrogenless models, the envelope is ionized by the shock radiation and the velocity at the photospheric level is already several thousand kilometers per second.

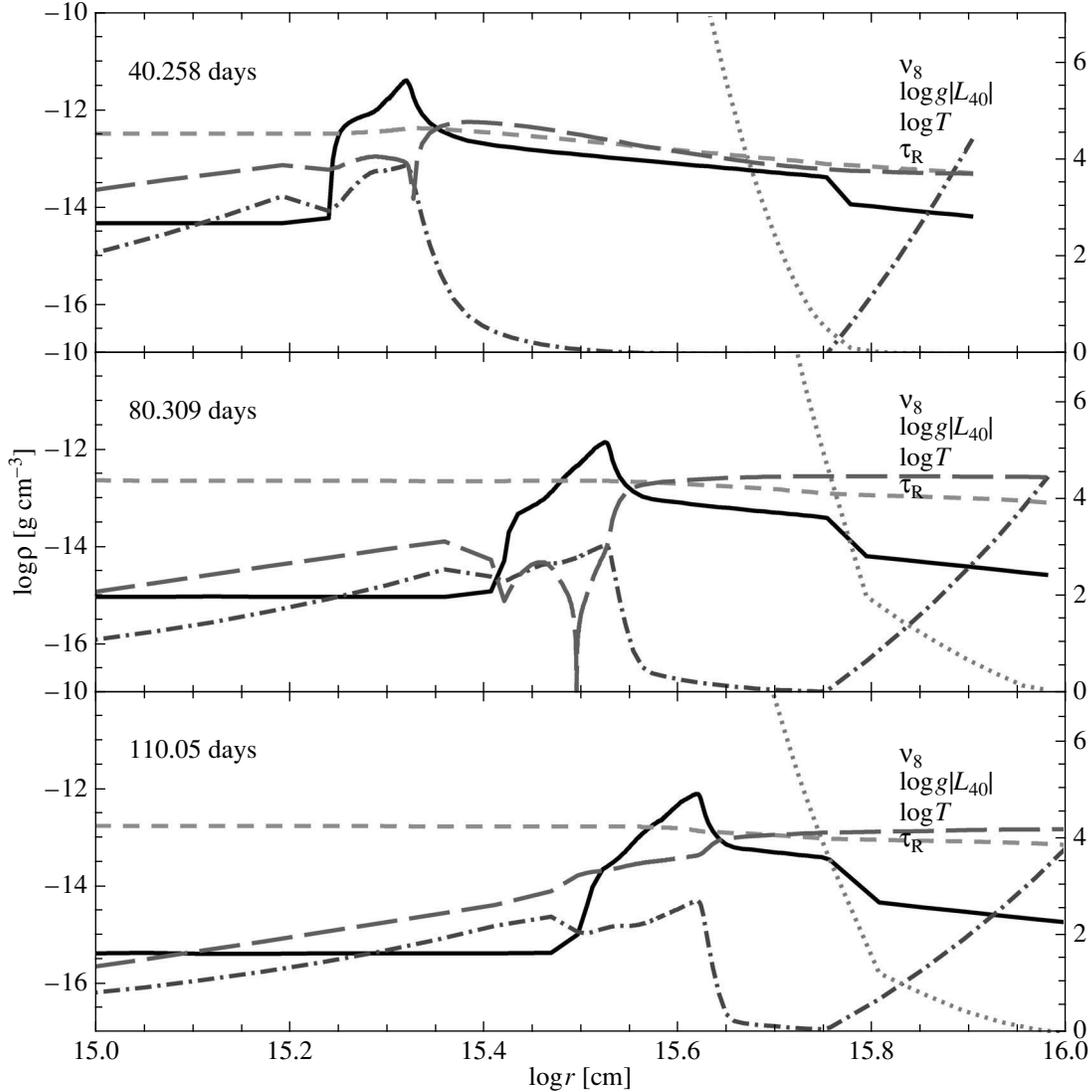

**Fig. 5.** Density $\rho$ (solid line), Rosseland optical depth $\tau$ (dotted line), temperature $T$ (short dashed line), luminosity $L$ (long dashed line), and velocity $v$ (dot-dashed line) versus radius $r$ for the helium `M53He48e40` model at the times $t = 40$, $80$, $110$ days. The logarithm of SN radius $r$ is along the horizontal axis. $\log(\rho)$ is along the vertical axis (left); $\log(\tau)$, $\log(T)$, and the normalized $\log(L_{40})$, $\log(v_8)$, where $L_{40} \equiv L/10^{40}$ erg s$^{-1}$ and $v_8 \equiv v/10^8$ cm s$^{-1}$, are along the vertical axis (right).

## CONCLUSIONS

Based on detailed radiation-hydrodynamics calculations, we constructed models for the SLSN PTF12dam. This object is interesting not only in that its luminosity at maximum light is enormous but also in that its light curve is very long. In addition, it was discovered at the rising phase of luminosity. As was shown by Nicholl et al. (2013), a comparatively fast flux rise is incompatible with the PISN (Pair-Instability Supernovae) models proposed previously for such SLSNe. Our modeling showed that all of the main characteristics (the rise time, the peak luminosity, the decay time, and the color temperature behavior) could be reproduced satisfactorily with a minimum set of model parameters and at a modest (for such superluminous objects) energy, ∼4 foe. Of course, some discrepancies in the behavior of the fluxes in different bands remain, but we doubt that it makes sense to further fit the parameters within the framework of our models, where the assumption about spherical symmetry is adopted, the possible clumpiness of the environment is ignored, and a local thermodynamic equilibrium in the properties of the material is assumed when the absorption and emission coefficients are calculated. The most important fact is that the global characteristics are reproduced even within the framework of such a simplified model.

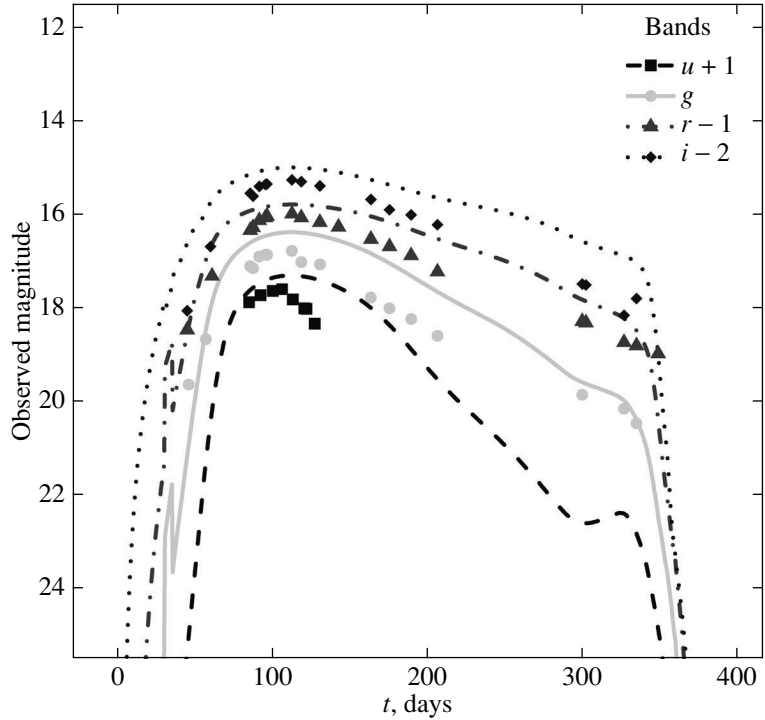

**Fig. 6.** *ugri* light curves for the carbon−oxygen model (C : O = 9 : 1) with a large total mass, $M = 100\,M_\odot$. The time in days from the explosion is along the horizontal axis, and the observed magnitude is along the vertical axis. The explosion time is MJD = 55986.

New observational data on the light curves for PTF12dam have been published recently (Chen et al. 2014). The power-law decline in the luminosity of PTF12dam continues at least for +400 days after the peak in the r band. To explain this behavior of the light curves, Nicholl et al. (2014) had to revise the magnetar model from their previous paper (Nicholl et al. 2013), which gives overestimated values for the late points on the bolometric light curve. They took into account the transfer of $\gamma$ radiation in the ejecta and obtained a new evaluative magnetar model that might explain the observed light curve.

The model with a shock also allows the observed power-law decline to be explained. We are planning to study PTF12dam in more detail with new observational data. Our preliminary modeling shows that an increase in the mass of the circumstellar envelope to $M = 100\,M_\odot$ leads to good agreement in the light curves not only for the bolometric light curve but also in various spectral ranges (see Fig. 6). In this case, we considered a carbon−oxygen envelope with C : O = 9 : 1, which is required to obtain the correct relationship between the peak fluxes in various spectral bands. The slope of the light curve after the peak coincides with the observed one better than in the model with helium.

Note that in all of the models proposed for SLSNe, one has to deal with large masses of helium stellar cores (occasionally more than 100 $M_\odot$) and carbon−oxygen cores (several tens of $M_\odot$). For example,

Moriya et al. (2010) proposed a model of a core-collapse supernova in which a carbon—oxygen core with a mass of 43 $M_\odot$ explodes instead of PISN. There is no envelope surrounding the exploding star and no shock in this model. The luminosity rises approximately twice as fast as it does in the PISN model, but this is achieved through the fact that an enormous explosion energy of 36 foe was specified "manually." An energy that is an order of magnitude lower is required in our models with a shock in a circumstellar envelope, which makes such models very attractive.

At a large explosion energy, Moriya et al. (2010) also obtain a very large amount of radioactive $^{56}$Ni, about 6.1 $M_\odot$, which is quite comparable to PISN. Recall that the presence of radioactive matter is not required in our models with a shock in an extended circumstellar envelope. Of course, it is conceivable that a combination of mechanisms, for example, a collapse with a more modest energy and some amount of $^{56}$Ni and a shock in an envelope, actually operates in nature. Such a combination will most likely allow one to reduce the total mass in the shock model that is required to explain the very long light curves in SLSNe.

Understanding such unique SNe as PTF12dam is extremely important both for the development of a stellar evolution theory and for the possible applications of SLSNe in cosmological problems. Therefore, the investigation of such objects must continue.

## ACKNOWLEDGMENTS

We thank K. Nomoto and R. Quimby for the useful discussions. This study was supported by a grant from the Russian Science Foundation (project no. 14-12-00203).

*Translated by V. Astakhov*